\input harvmac
\noblackbox

\def\nup#1({Nucl.\ Phys.\ $\us {B#1}$\ (}
\def\plt#1({Phys.\ Lett.\ $\us  {B#1}$\ (}
\def\cmp#1({Comm.\ Math.\ Phys.\ $\us  {#1}$\ (}
\def\prp#1({Phys.\ Rep.\ $\us  {#1}$\ (}
\def\prl#1({Phys.\ Rev.\ Lett.\ $\us  {#1}$\ (}
\def\prv#1({Phys.\ Rev.\ $\us  {#1}$\ (}
\def\mpl#1({Mod.\ Phys.\ Let.\ $\us  {A#1}$\ (}
\def\ijmp#1({Int.\ J.\ Mod.\ Phys.\ $\us{A#1}$\ (}
\def\jag#1({Jour.\ Alg.\ Geom.\ $\us {#1}$\ (}
\def\tit#1|{{\it #1},\ }

\def\Coe#1.#2.{{#1\over #2}}
\def\coeff#1#2{\relax{\textstyle {#1 \over #2}}\displaystyle}
\def\coe#1.#2.{\relax{\textstyle {#1 \over #2}}\displaystyle}
\def\half{{1 \over 2}}

\def\del{\partial}

%
\lref\HKLM{J. Harvey, P. Kraus, F. Larsen and E. Martinec, JHEP 0007 (2000) 042,
{\it D-branes and strings as non-commutative solitons}, hep-th/0005031. }
\lref\EW{E.\ Witten,  Phys.\ Rev. D46 (1992) 5467, {\it On Background Independence of Open String Field Theory},
 hep-th/9208027.}
\lref\EWW{E.\ Witten,  Phys.\ Rev. D47 (1993) 3405, {\it Some Computations in Background Independent
 Open String Field Theory},
 hep-th/9210065.}
\lref\KMM{D.\ Kutasov, M. Marino and G. Moore , {\it Some Exact results on Tachyon Condensation,
} hep-th/0009148.}
\lref\SS{S. Shastashvili, Phys.\ Lett.\ B311 (1993) 83, {\it Comment on the Background Independent
 Open String Field Theory,}
 hep-th/9303143; {\it On the Problems with the Background Independent
 Open String Field Theory,}  hep-th/9311177.}
\lref\GS{A. Gerasimov and S. Shastashvili,{\it On  Exact Tachyon Potential in 
 Open String Field Theory,}
 hep-th/0009148. }
\lref\ACNY{A. Abouelsaood ,C. Callan , C. Nappi and S. Yost, Nucl.\ Phys.\ B280 (1987) 599{\it 
Open Strings in Background Gauge Fields.}
  }
\lref\KO{K. Okuyama, {\it 
Noncommutative Tachyon from Background Independent Open String Field Theory,} hep-th/0010028. }
\lref\KMMM{D.\ Kutasov, M. Marino and G. Moore,{\it Remarks on Tachyon Condensation in 
Superstring Field Theory, }hep-th/0010108. }
\lref\DMR{K. Dasgupta, S. Mukhi and G. Rajesh, JHEP 0006 (2000) 022,{ \it
Noncommutative tachyons}, hep-th/0005006.}
\lref\WN{E. Witten, {\it Noncommutative tachyons and string field
theory}, hep-th/0006071.}
\lref\SW{N. Seiberg and E. Witten, JHEP 9909 (1999) 032, {\it String Theory and
Noncommutative Geometry}, hep-th/9908142.}
\lref\AT{A. Tseytlin, {\it Sigma Model Approach to String Theory 
Effective Actions with Tachyons,} hep-th/0011033.}
\lref\OA{O. Andreev, {\it Some Computations of Partition Functions and Tachyon Potentials in Background Independent 
Off-Shell String Theory,}hep-th/0010218.}

\Title{\vbox{
\hbox{CITUSC/00-051}
\hbox{USC-00/06}
\hbox{\tt hep-th/0011108}
}}{\vbox{\centerline{\hbox{{Background Independent Open  String 
}}}
\vskip 8 pt
\centerline{ \hbox{Field Theory and Constant $B$-Field }}}}
\centerline{ D.~Nemeschansky 
and V.Yasnov}
\bigskip
\centerline{\it Department of Physics and Astronomy}
\centerline {and}
\centerline {CIT-USC Center of Theoretical Physics }
\centerline{University of Southern California}
\bigskip

\vskip .3in

We calculate the background independent action for bosonic and supersymmetric 
open string field theory 
in a constant $B$-field. We also determine the tachyon effective action 
in the presence of constant $B$-field.

\vskip .3in



%
\parskip=4pt plus 15pt minus 1pt
\baselineskip=15pt plus 2pt minus 1pt
%
 %
Recently there has been  a renewed interest in Witten's \EW ,\EWW\ background independent formulation
of open string field 
theory    due to its relation to the tachyonic lagrangians. In \GS\ and \KMM\ the tree
level lagrangian was constructed from the background independent open string field theory. 
The question of tachyon condensation in the presence of large $B$-fields, which
is equivalent
to large world volume gauge fields\DMR,\WN , seem to lead to great
simplifications. The non-commutative geometry \SW\ simplifies the construction
of soliton solutions which correspond to $D$-branes \HKLM .

In this letter we  generalize the analysis of \EWW ,\GS\ and \KMM\ to the case of 
constant $B$-field.
Following Witten's construction we consider a
two-dimensional action of the form 
\eqn\act{I\ = \ I_0+I^\prime \ \ ,}
where $I_0$ is the bulk action and the boundary term $I^\prime$ describes coupling of 
the open strings. 
In this letter the bulk action describes the closed string background with a
constant B-field
\eqn\bulk{ I_0 \ = \ \int_\Sigma  d^2 x \sqrt h \Bigl ( {1 \over 8 \pi }
h^{ij} \del_i X^\mu \del_j X_\mu +b^{ij} D_i c_j \Bigr) -{ i \over 8 \pi}
\int_\Sigma  d^2 x \epsilon^{ij} B_{\mu \nu} \del_i X^\mu \del_j X^\nu 
 \ \ . }
Here $\Sigma$ is the two-dimensional world sheet with the metric $h_{ij}$ and
coordinates $X^i$. The ghost and anti-ghost fields are given by $c_i$ and
$b^{ij}$. 
The new feature of our analysis is to include a constant  background $B$-field in the action.
The theory is $ BRST $ invariant and has associated with it a conserved current
$J^i$.
The corresponding charge $Q= \oint d \sigma J^0$ satisfies 
\eqn\brst{ Q^2 \ = \ 0 \ \ \ \ \ T_{ij} \ = \ \{Q , b_{ij} \} \ \ .}
Above $T_{ij}$ is the stress energy tensor of the two-dimensional theory. The
boundary interaction has the form
\eqn\bound{ I^\prime \ = \ \int_{\del \Sigma} {\cal V} \ \ ,}
where the operator $\cal V$ is a local operator constructed out of the fields $X,c$
and $b$.
Witten \EW ,\EWW\ has defined a gauge invariant Lagrangian on the space of open string
field theories with the action \act . 
We use  this approach to determine the invariant Lagrangian in the
presence of a constant $B$-field.

The ghost number of the operator $\cal V$ is zero, but it still may depend on
the ghost fields. 
Following the construction of \EWW\ one must also introduce a ghost number one
operator $\cal O $ via
\eqn\callo{{ \cal V} \ = \ b_{-1} \cal O \ \ .}
For the bosonic theory $\cal V$ can be constructed purely out of the matter
fields.
In a derivative expansion it has the form
\eqn\derex{{\cal V} \ = T(x) +A_\mu (x) \partial_\sigma X^\mu +
C_{\mu\nu}\partial_\sigma X^\mu \partial_\sigma X^\nu +\cdots \ \ .} 
Since  only  matter fields appear in Eq. \derex\ the ghost number one operator $\cal O $ is given by
\eqn\newcalo{ {\cal O} \ = \ c {\cal V}\ \ .}
The action $S$ of the background independent open string field theory is a
functional of the boundary perturbation.
It can be determined from the following bulk correlation function
\eqn\defs{ d S \ = \ < d \int_{\del \Sigma} {\cal O} \{Q , \int_{\del \Sigma} 
{\cal O} > \} \ \ . }
In the BV formalism this can be written as 
\eqn\bvform{ d S \ = \ i_V \omega\ \ \ ,}
where $\omega$ is an odd sympletic structure and $V$ is a vector field that
generates the symmetries of $\omega$.

In \EWW\ it was shown that the action $S$ is related to the partition function $Z$
by the following equation
\eqn\relsz{ S \ = \ (  - \beta^i{ \del\over \del u^i} +1 ) Z \ \ \ ,
 }
where $\beta^i$ is the beta-function for the coupling $u^i$.
At the fixed point $u^\star_i\ = \ 0$
\eqn\fixpoint{S(u^\star_i )\ = \ Z(u^\star_i)\ \ .}

In this letter we consider a quadratic boundary perturbation of the form
\eqn\quadper{{\cal O} \ = \ cT \ \ \ \ \ {\it and}\ \ \ \ \  T \ = \sum {u_\mu \over 8 \pi
}(X^\mu)^2\ \ \ . }
The action describing the scalar fields $X^\mu $ is given by
\eqn\quadac{\eqalign{S \  & ={ 1 \over 8 \pi} \int _{\Sigma} \Bigl(\Bigl( {\del X^\mu \over
\del \sigma^1}\Bigr )^2 +\Bigl ( {\del X^\mu \over \del \sigma^1}\Bigr )^2 - i B_{\mu\nu}{\del
X^\mu \over \del \sigma^1}{\del X^\nu \over \del \sigma^2}\Bigr )d^2 \sigma+\cr & {i \over 8 \pi
}\int_{\del\Sigma}d\theta u_\mu (X^\mu)^2 \ \ \ .}}
Since we are interested in calculating the tree level action we take as our world sheet 
a disk $D$ with a rotationally invariant metric.
The flat metric on the disk is given by
\eqn\flatmetr{ ds^2 \ = \ d\sigma_1^2 + d \sigma_2^2\ \ \ \ \ \ \
\sigma^2_1+\sigma^2_2\ \le \ 1\ \ .}
It is convenient to work with complex coordinates and we set $z=\sigma_1+i
\sigma_2$.
The boundary of the disk in these coordinates is at $|z|\ = \ 1$ and we parametrize the boundary by
$z= e^{i \theta}$.
Varying the action \quadac\ we find that in the presence of the boundary term
the scalar fields $X^\mu$ must satisfy the following boundary conditions
\eqn\boundarycond{(1+B)_{\mu\nu} z \del X^\nu +(1-B)_{\mu\nu}\bar z \bar \del
X^\nu +u_\mu X^\mu |_{\del \Sigma} \  = 0 \ \ .
}

To  determine the action we need  
Green function  $G(z,w)$ of the theory.
It  satisfies the following equation 
\eqn\greenf{\del_z\bar\del_{\bar z} G(z,w) \ = \ - 2 \pi \delta^{(2)}(z-w) \ \
\ }
along with the boundary condition \boundarycond . 
The Green function for the case with no boundary term ($u=0$) has been constructed in \ACNY\
and it has  been used in the analysis of non-commutative geometry.
Extra terms have to be added  to this Green function in order to satisfy the boundary conditions \boundarycond .
The Green function with the boundary term is given by
\eqn\solg{G(z,\omega)\ = \ - ln |z-\omega|^2-{1-B \over1+B} ln
(1-z\bar{\omega})-{1+B \over 1-B} ln
(1-\bar z \omega)+{2 \over u }-2f_1(z\bar{\omega})-2f_2(\bar{z}\omega) \ \ ,}
where 
\eqn\deff{\eqalign{ & f_1(x)
\ = \ {1 \over 1+B}\sum_{k=1}^\infty {1\over k}{u \over 1+B}  \Bigl (k+{u\over 
1+B} \Bigr)^{-1} x^k \cr &
f_2(x)
\ = \ {1 \over 1-B}\sum_{k=1}^\infty {1\over k}{u \over 1-B}  \Bigl (k+{u\over 
1-B} \Bigr)^{-1} x^k \ \ \ . \cr} 
}
To verify that the Greens function \solg\ satisfy the correct boundary
conditions it is useful to note that the functions $ f_1(z)$ and $ f_2(z)$ 
satisfy the following identity
\eqn\identity{\eqalign{ &(1+B)z\partial
f_1(z\bar{\omega})+uf_1(z\bar{\omega}) \ = \ - {u\over 1+B} ln (1-z\bar
{\omega}) \cr &(1-B)z\partial
f_2(z\bar{\omega})+uf_1(z\bar{\omega}) \ = \ - {u\over 1-B} ln (1-z\bar
{\omega}) \ \ .}}

The first step towards determining the effective action is to calculate the
partition function $Z(u)$ on the disk,
\eqn\diskpart{ Z(u) \ = \ \int {\cal D} X \ exp(-I) \ \ .}
To simplify our presentation we consider the case where the only non-zero
component of the $B$-field is $B_{12}\ = -B_{21} \ = \ b$. The result can be
generalized easily as we see below.
In this case we have two scalar fields $X^1$ and $X^2$ with $u_1=u_2=u$. The
boundary term is given by
\eqn\speboun{ {1 \over 8 \pi } \int_{\del \Sigma} d \theta u \Bigl(  X_1^2(\theta )+ X_2^2(\theta) \Bigr)
\ \ . }
We need to define quantum operators $X_i^2(\theta)$ appearing in \speboun\ .
Using point splitting  we define 
\eqn\qunt{X_i^2(\theta )\ = \ \lim_{\epsilon \rightarrow 0}(X_i(\theta)
X_i(\theta+\epsilon)-f(\epsilon )\ \  , }
where 
\eqn\defff{f(\epsilon) \ = \ - { 2 \over 1+B} ln(1- e^{i\epsilon})
 - { 2 \over 1-B } ln(1- e^{-i\epsilon}) \ \ .}
With the above definitions we have
\eqn\expect{\langle X_1 (\theta)X_1 (\theta)\rangle=\langle X_2 
(\theta)X_2 (\theta)\rangle \ = \
{2\over u}-{4u\over 1+b^2}\sum_{k=1}^\infty
{1\over k}{k+u-b^2k\over {(k+u)}^2+b^2k^2}\ \ .}
Now we are ready to  calculate the partition function on the disk. 
We have
\eqn\derpart{\eqalign{{d\over du} ln Z \  & = \ - {1\over 8\pi}\int_0^{2\pi}d\theta
\Bigl (\langle X_1 (\theta )X_1 (\theta)\rangle +\langle X_2 (\theta)X_2
(\theta)\rangle\Bigr ) \ \cr & =
-{1\over u}+{2 u\over
1+b^2}\sum^\infty_{k=1}{1\over k}{k+u-b^2k\over {(k+u)} ^2+b^2k^2}\cr } }
In order to integrate for the partition function it is useful to rewrite the right hand
side in \derpart\ as follows
\eqn\rewrite{\eqalign{{d\over du} ln Z \ & = \ -  {1 \over u}+{1\over 1-ib}\sum^\infty_{k=1}
{1\over k}{{ u\over (1-ib)} \over k+{u\over (1-ib)}}\cr &
+{1\over 1+ib}\sum^\infty_{k=1}
{1\over k}{{ u\over (1+ib)} \over k+{u\over (1+ib)}} \ \ \ .}}
The two sums appearing in \rewrite\ are related to Euler function $\Gamma(u)$,
\eqn\defgamma{ \Gamma(u) \ = \ {e^{-\gamma u}\over u }
\prod_{n=1}^{\infty}(1+{u\over n})^{-1}e^{u\over n}\ \ ,
}
where the constant $\gamma$ is called Euler constant
\eqn\defeuler{\gamma \ = \ \lim_{n \rightarrow \infty} \ (1+ \coeff{1}{2}+ \coeff{1}{3}+ \ldots + \coeff{1}{n}-\ln
\ n )\ \ \ .}
The gamma functions has the following property
\eqn\gammaprop{{d \over du} ln \Gamma(u) \ = \ - { 1 \over u} + \sum_{n=1}^\infty { u \over 
k(k+u)}- \gamma \ \ .}
Using this formula we can write \rewrite\ in the form
\eqn\regamma{{d\over du} \ln Z(u) \ = \ {1\over u}+{2 \gamma\over 1+b^2}+{d \over du} \ln\Gamma
({u\over 1+ib})+{d \over du} \ln\Gamma
({u\over 1-ib})\ \ \ .}
Now it is an elementary exercise to determine the partition function. 
We find up to a $u$ independent normalization constant
\eqn\finalpartion{Z(u) \ = \ u \ exp \Bigl( {2\gamma u\over 1+b^2}\Bigr )\Gamma
\Bigl({u\over 1+ib}\Bigr) \Gamma \Bigl( {u\over 1-ib}\Bigr )\ \ .}
Having determined the partition function for the special case of a single non-vanishing
component of the $B$-field it is easy to determine the partition function for the general
case. 
Note  that the eigenvalues of the matrix $ \coeff{u}{1-B^2}$ are $\coeff{u}{1+i b}$ and 
$\coeff{u}{1-i b}$.
Using this we then have
\eqn\detgamma{{ \it  \det} \ \Gamma \Bigl ({u\over 1+B}\Bigr )\  = \ \Gamma
\Bigl({u\over 1+ib}\Bigr)\Gamma \Bigl ({u\over 1-ib}\Bigr )\ \ .}
To generalize the term that multiplies Euler's constant we note that
\eqn\euler{{\it \Tr} { 1\over 1- B^2} \ = \ {2 \over 1+b^2}}
Collecting all the terms we have that the partition function has the form
\eqn\generalpart{ Z(u) \ = \ u \ {\it \exp }\Bigl (\gamma \rm{Tr}{u\over 1-B^2}\Bigr
)\sqrt{{ \det} \Gamma \Bigl ( {u\over1+B}\Bigr )\Gamma \Bigl ({u\over 1-B}\Bigr )}}

Having determined the partition function, we can calculate the background independent action
from
\eqn\wittenact{dS \ = \ {1\over 2}\int^{2\pi}_0d\theta d\theta '\langle d{\cal O}(\theta)\
\{Q,{\cal
 O}\}(\theta ')\rangle \ \ ,}
where
\eqn\defcaloo{{\cal O} \ = \ { 1\over 8\pi}c \ u(X_1^2+X_2^2)\ \ .} 
As before, we first  carry out our calculation for the special case of $B_{12} =  -B_{21}
= b$.
A simple calculation shows that in the presence of a constant $B$-field
\eqn\commut{ \{Q,{\cal O}\} \ = \ 
{1\over 8\pi} cc'\Bigl ( \sum_{i=1}^2 {1\over 1+b^2}{\partial^2\over \partial
X_i^2}+1\Bigr)
(X_1^2+X_2^2)\ \ . } 
The ghost correlator that we need in the background independent action is $< c(\theta) c
c^\prime(\theta^\prime)>$.
Using the fact that three ghost zero modes on  the disk are $1$, $e^{i\theta}$ and
$e^{-i\theta}$, and normalizing the ghost measure so that $<c c^\prime
c^{\prime\prime}(\theta) >\ = \ 1$ %
\eqn\ghostcorr{ < c(\theta) c c^\prime(\theta^\prime) > \ = \ 2 (cos(\theta-
\theta^\prime)-1) \ \ .}
Consequently the  defining equation for $S$ has the form
\eqn\defings{\eqalign{d S \ = & \half \int^{2 \pi}_0 d \theta \ d\theta^\prime \ 2  ( cos(\theta
\ -\theta^\prime)-1)\ \cr &\Bigl < \Bigl ( 
u{X_1^2(\theta)+X_2^2(\theta)\over 8 \pi}\Bigr) { 1 \over 8 \pi} \Bigl( { 4u \over
1-b^2}+ u (X_1^2(\theta^\prime)+X_2^2(\theta^\prime)\Bigr)\Bigr >}}
To determine $S$ we need to evaluate the following correlation function
\eqn\corr{W \ = \int^{2\pi}_0 {d\theta d\theta^\prime \over{(2\pi)}^2}\cos
(\theta-\theta^\prime) \langle (X_1^2+X_2^2)(\theta)(X_1^2+X_2^2)(\theta ')\rangle \ \ .
 }  
Using the explicit form of the Green's function on the boundary
\eqn\boundarygreen{\eqalign{&
G_{11}(\theta,\theta^\prime)  =2\sum_{-\infty}^{\infty}e^{ik(\theta - \theta^\prime)}{|
k|+u\over{(|k|+u)}^2+b^2k^2} \cr & G_{12}(\theta,\theta^\prime)\ = \
-2\sum_{-\infty}^{\infty}e^{ik(\theta -\theta ')}{b k\over{(|k|+u)}^2+b^2k^2}\ \ .}}
and  doing the angular integrations we find
\eqn\integr{W \ = \ 8\Bigl
({1\over {(1-ib)}^2}\sum_{-\infty}^{\infty}{1\over |k|+u_-}{1\over| k+1|+u_-}+
{1\over {(1+ib)}^2}\sum_{-\infty}^{\infty}{1\over|k|+u_+}{1\over|k+1|+u_+}\Bigr )\ \ , }
where
\eqn\defup{\eqalign{u_- \ = \ {u\over 1-ib}&\cr u_+\ = \ {u\over 1+ib} \ \ \ .}}
Most of the terms cancel in the sum \integr\ and we are left with
\eqn\resW{W\ = \ {32\over u}{1\over 1+b^2 \ .}}
The rest of the correlators that appear in \defings\ can be evaluated, using
\eqn\restcorr{\eqalign{ \int^{2 \pi}_0 d \theta <X_1^2(\theta)+X_2^2(\theta)> = -8\pi {d
Z(u) \over d u}\cr \int^{2 \pi}_0 d \theta \int^{2 \pi}_0 d \theta^\prime <(X_1^2(\theta)+X_2^2(\theta))(X_1^2(\theta^\prime)+X_2^2(\theta^\prime))> 
= (8\pi )^2{d^2 Z(u)
\over d^2 u }\ \ .}} 
Using \resW\ and \restcorr\  we have
\eqn\finalds{dS=\ {2\over 1+b^2}Zdu+{2 \over 1+b^2}{dZ\over
 du}udu-u{d^2Z\over du^2}du \ \ .}
Integrating this we get
\eqn\ress{S=\Bigl ( {2u \over 1+b^2}+1-u{d\over du}\Bigr )Z\ \ .}
As before we can again generalize this for an arbitrary constant $B$-filed background.
Using the identity  $ \coeff{2u}{1+b^2}= {\rm Tr} \coeff{u}{1-B^2}$ we find that in the
general case
\eqn\generals{S\ = \ {\rm Tr}\Bigl ( {u\over 1-B^2}+1-u {d\over du} \Bigr)Z\ \ .}
This is our final result for the general form of the space-time action for boundary interactions 
we have considered.

Next we use this to determine effective action for the tachyon field by expanding around $u=0$.
As we have done through out this paper we first consider the case where the only non-vanishing component of the $B$-field is 
$B_{12} =  -B_{21} =  b$. 
Expanding the partition function  around $u=0$  we have
\eqn\partuzero{Z \  = \ {1+b^2\over u}+O(u^2)\ \ \ .}
Rewriting this in terms of the tachyon field $T(X)=-u(X_1^2+X_2^2)$ we get
\eqn\usetach{Z\ = \ {1\over \pi}\int dX_1dX_2{(1+b^2)}^{1/2}e^{-T(X)}\ \ .}
To get the partition function for the general case one has to replace $1+b^2$ by
 $det(1+B) $ since  $det(1+B) \ = \ 1+b^2$.
 
So far we have not fixed the overall normalization of the partition function.
As we mentioned earlier  we are free to multiply the partition function by any $u$-independent function.
The $B$-dependence can be fixed by comparing with the Born-Infeld action. 
We find that our partition function for the general case has to be multiplied by a factor
$\det^{-1/2} (1+B)$.
Changing the normalization of \usetach\ we have
\eqn\newzfinal{Z\ = \ {1\over \pi}\int dX_1dX_2{(1+b^2)}^{1/2} e^{-T(X)}\ \ .}

To find the effective  action for the tachyon field we calculate the open string
effective action $S$ using the partition function \newzfinal ,
\eqn\finalsform{S\ = \ {1\over \pi}\int dX_1dX_2{(1+b^2)^{\half}}\Bigl({2 u \over 1+b^2} +1 - u {d \over du}\Bigr )
e^{-T(X)} \ \ .}
To evaluate  \finalsform\ we use
\eqn\ident{\eqalign{\Bigl  ({2u\over 1+b^2}+1\Bigr ) e^{-T(X)}\ & =\Bigl ( 1+{1\over 2(1+b^2)}
(\partial_1^2 T(X)+\partial_2^2 T(X))\Bigr ) e^{-T(X)}\cr
-u{d \over du}e^{-T(X)}\ & = \ T(X)e^{-T(X)}\ \ .}}
After substituting \ident\ into  \finalsform\ and integrating by parts we obtain 
the action for the tachyon field
\eqn\finaltachyonact{S \ = \ {1\over \pi}\int dX_1dX_2\sqrt{1+b^2}e^{-T(X)}\Bigl (
{1 \over 2(1+b^2)}(\partial_1 T(X) \partial_1 T(X)+\partial_2 T(X) \partial_2 T(X))+1+T\Bigr )\ \ \ . }
To generalize this to an arbitrary constant $B$-field we note that
\eqn\identi{ {1 \over 1+b^2} \ = \  { 1 \over (1-B^2)_{11}}\ = \  { 1 \over (1-B^2)_{22}}\ \ .}
Using \identi\ we find that the tachyon action for the general constant background field  $B$ is of the form
\eqn\general{S \ = \ {1\over \pi^{n/2}}\int d^nX \sqrt{\det (1+B)}e^{-T(X)}\Bigl ({1 \over2}
{\Bigl ({1\over 1-B^2}\Bigr )}_{\mu\nu}\partial_\mu T(X)\partial_\nu T(X)+1
+T(X)\Bigr )\ \ .}

The results of the present paper can be generalized a supersymmetric model \KMMM,\AT,\OA\  with non-vanishing $B$-field.
The boundary interaction is 
\eqn\susybound{ S_{boundary}\ = \coeff{1}{8\pi}u\int^{2\pi}_0 d\theta\sum_{i=1}^2 \left ( X_i^2+\coeff{1}{2}(\psi_i{1\over \partial_\theta}\psi_i+
\bar{\psi}_i{1 \over \partial_\theta}\bar{\psi}_i)\right ) \ . }
To compute the partition function with the boundary action \susybound\ we need the fermionic 
Green's functions in the presence of a $B$-field. In the $NS$-sector we have
\eqn\susygreen{\eqalign{ G_F(\theta-\theta^\prime) \ &  \equiv \ \langle\psi (\theta)\psi 
(\theta^\prime)\rangle = 2i\sum_{r=Z+1/2} {r\over (1+B)|r|+u}e^{ir(\theta-\theta^\prime)} \cr
\bar G_F(\theta-\theta^\prime) \ &  \equiv \ \langle\bar \psi (\theta)\bar \psi 
(\theta^\prime)\rangle = 2i\sum_{r=Z+1/2} {r\over (1-B)|r|+u}e^{ir(\theta-\theta^\prime)} \ \ .\cr
}}
In addition to these Green's functions we also need  the following Green's functions
\eqn\addgreen{\eqalign{g_F(\theta-\theta^\prime) \ & \equiv \ \langle\psi (\theta)
{1\ \over \partial_{\theta^\prime} }\psi (\theta^\prime)\rangle \  = \
-2\sum_{r=Z+1/2}
 {1\over (1+B)|r|+u}e^{ir(\theta-\theta^\prime)} \cr
\bar g_F(\theta-\theta^\prime)\ & \equiv \ \langle\bar \psi (\theta)
{1\over \partial_{\theta\prime  }}\bar \psi (\theta^\prime)\rangle \  = \
-2\sum_{r=Z+1/2} {1\ \over (1-B)|r|+u}e^{ir(\theta-\theta^\prime)}\ \ . \cr }}
The partition function can be determined from the correlator
\eqn\corr{\sum_{i=1}^2 \langle  X_i^2+\coeff{1}{2}(\psi_i{1\over \partial_\theta}\psi_i+
\bar{\psi}_i{1 \over \partial_\theta}\bar{\psi}_i)\rangle\ \ . }
To define this correlator we use the point splitting method
\eqn\corrsplit{\eqalign{&\sum_{i=1}^2 \langle  X_i^2+\coeff{1}{2}(\psi_i{1\over \partial_\theta}\psi_i+
\bar{\psi}_i{1 \over \partial_\theta}\bar{\psi}_i)\rangle \  \ = \cr & \lim_{\epsilon \rightarrow 0}
\sum_{i=1}^2 \langle  X_i(\theta+\epsilon) X_i(\theta)+\coeff{1}{2}(\psi_i(\theta+\epsilon) {1\over \partial_\theta}\psi_i(\theta)+
\bar{\psi}_i(\theta+\epsilon){1 \over \partial_\theta}\bar{\psi}_i(\theta))\rangle \
= \cr & \lim_{\epsilon \rightarrow 0} (2 G_B(\epsilon) + g_F(\epsilon)+ \bar g_F(\epsilon) ) \ \ .}}
As before,  the bosonic Green's function can be written as
\eqn\bosg{\eqalign{G_B(\epsilon )= & -{2\over 1+B}f(u_+,\theta )-{2\over 1-B}f(u_-,\theta )-\cr
& {2\over 1+B} \ln (1-e^{i\theta })-
{2\over 1-B}\ln (1-e^{-i\theta }) +\coeff{2}{u}\ \ , \cr }}
where $u_+ = u/(1+B)$ , $u_-= 1/(1-B)$ and 
\eqn\defifu{f(u,\theta)\equiv\sum^\infty_{k=1}{u\over k(k+u)}e^{ik\theta\ \ .}}
To evaluate the fermionic Greens function in this limit  we first note that 
the index $k$ in Eq. \addgreen\ runs over one half of an odd number.
One can rewrite the sum over odd integers as a difference of all integers and even integers.
This result in two sums that are similar to the bosonic Green function and hence can be written in terms of the function $f(u, \theta)$:
\eqn\gepsilon{\eqalign{
g_F(\epsilon) \ & =
 \ {1 \over 1+B }\bigl (\ln|1-e^{i\theta/2}|^2-2\ln|1-e^{i\theta}|^2+8f(2u_+,\theta)-4f(u_+,\theta)\bigr ) \cr
\bar{g}_F (\epsilon) \ & ={ 1 \over 1+B} 
\bigl
 ( 4\ln {|1-e^{i\theta/2}|}^2-2\ln{|1-e^{i\theta}|}^2+8f(2u_-,\theta)-4f(u_-,\theta)\bigr ) \ \ . \cr }}
In the supersymmetric case the combination of Green's functions
\eqn\combo{\lim_{\epsilon \rightarrow 0} (2 G_B(\epsilon) + g_F(\epsilon)+ \bar g_F(\epsilon) )}
is finite so that no regularization is needed.
Using the Green's functions constructed above we find
\eqn\derz{\eqalign{{d\ln Z\over du}= & -\coeff{1}{u}+{\rm Tr}\bigl ({1\over 1+B}(f(u_+,0)-f(2u_+,0))+{1\over 1-B}(f(u_-,0)-f(2u_-,0))\bigr )-\cr
& {1\over 1-B^2}2\ln 4 \cr}}
To integrate for the partition function we recall that the function $f(u,0)$ can be writen in terms of derivatives of Gamma functions.
After some simple algebra we find
\eqn\partz{Z \ = \ u \det {\Gamma (u_+)\Gamma (u_-)\over \sqrt{\Gamma (2u_+)\Gamma (2u_-)} } \exp {\rm Tr} \bigl (  {u\ln 4\over 1-B^2}\bigr)}
The effective action for the tachyon field in the supersymmetric case can be obtained by expanding the partition function around $u=0$.
We carry out this expansion first for the case of $B_{12}=b=-B_{21}$. 
To get the correct normalization for the tachyon potential one has to multiple the effective action $Z$ 
 by $\det^{-1/2} (1+B)$, which results in
\eqn\effu{Z(u)={4(1+b^2)^{1/2}\over u}\bigl ( 1+{2u\over 1+b^2}\ln 4\bigr)\ \ .} 
Rewriting this in terms of the tachyon field we have
\eqn\tachsusyef{S\ = \ \coeff{4}{\pi}\int dx_1dx_2 {(1+b^2)}^{1/2}e^{-T(x)}\bigl (1+{\ln 2\over 1+b^2}(({\partial_1 T(x)})^2+({\partial_2 T(x)})^2)\bigr )\ \ .}
The effective action for the general case can be easily deduced and we find
\eqn\finalfor{S \ = \ \coeff{4}{\pi^{n/2}}\int d^nx \ {\det}^{1/2} (1+B)e^{-T(x)}\bigl ( 1+\ln 2\bigl ( {1\over 1-B^2}\bigr )_{\mu\nu}\partial_\mu
 T(x)\partial_\nu T(x)\bigr )\ \ . }
 In the supersymmetric case we have used the  assumption \KMM\ that the effective action is given by partition function.
Justification of this assumption requires further investigations.
Clearly one would like to have a generalization of the Eq. \defs\ to the supersymmetric case.
One would also like to verify that the equation of motion that arise from 
the effective action vanish when the two-dimensional theory with the boundary interaction is conformal \SS ,\GS.

It would be 
 interesting to try and generalize the calculations of the effective action beyond tree level results by including
the annulus diagram. 
In particular it would be interesting to see the form of the exponential in the tachyon potential in higher order terms, since at 
 the tree the tachyon field  seems to be playing  the role of the loop counting parameter.

\bigskip
  
{\bf Acknowledgments} We would like to thank S. Shastashvili for discussions.
This work was supported in part by fund provided by the DOE under grant  number DE-FG03-84ER-40168

\bigskip
  
{\bf Note added} While completing this work, we learned that similar results to present paper has been obtained independently in \KO .
\goodbreak 
\listrefs

\vfill
\eject
\end